# The connection between starspots and superflares: a case study of two stars

Alexandre Araújo,[1]⋆ Adriana Valio,[1]

[1]*Center for Radio Astronomy and Astrophysics Mackenzie, Mackenzie Presbyterian University*
*Rua da Consolação, 860, São Paulo, SP - Brazil*

## ABSTRACT

How do the characteristics of starspots influence the triggering of stellar flares? Here we investigate the activity of two K-type stars, similar in every way from mass to rotation periods and planetary systems. Both stars exhibit about a hundred spots, however Kepler-411 produced 65 superflares, while Kepler-210 presented none. The spots of both stars were characterized using the planetary transit mapping technique, which yields the intensity, temperature, and radius of starspots. The average radius was $(17 \pm 7) \times 10^3$ km and $(58 \pm 23) \times 10^3$ km, while the intensity ratio with respect to the photosphere was $(0.35 \pm 0.24)$ $I_c$ and $(0.64 \pm 0.15)$ $I_c$, and the temperature was $(3800 \pm 700)$ K and $(4180 \pm 240)$ K for spots of Kepler-411 and Kepler-210, respectively. Therefore, spots on the star with no superflares, Kepler-210, are mostly larger, less dark, and warmer than those on the flaring star, Kepler-411. This may be an indication of magnetic fields with smaller magnitude and complexity of the spots on Kepler-210 when compared to those on Kepler-411. Thus, starspot area appears not to be the main culprit of superflares triggering. Perhaps the magnetic complexity of active regions is more important.

## 1 INTRODUCTION

Solar flares are transient phenomena that occur in the solar atmosphere in regions of high magnetic field concentrations, where a large amount of energy is released into the corona due to reconnection of magnetic field lines (Benz 2017). High precision photometric observation, such as that of the Kepler space mission (Koch et al. 2010), allowed detailed analysis of magnetic activity phenomena, such as starspots and faculae, superflares, and rotation of thousands of stars. The results further showed that stars with superflares exhibit an almost periodic brightness modulation, caused by the presence of large stellar spots on their surface. Thus, it is believed that flares in solar-type stars are also powered by magnetic reconnection.

Currently, the great challenge is unveiling the mechanisms that cause superflares. On the Sun, we know that the most energetic X-ray flares tend to occur on large sunspots (Sammis et al. 2000). During the last decades, a large number of studies have revealed the nature of the processes that may influence flare occurrence on various timescales, from the build up of energy to the triggering of flares. In their review, Toriumi & Wang (2019) cite as the main causes of solar flares the magnetic complexity, new flux emergence, shear motion, sunspot rotation, and magnetic helicity injection. The authors conclude that for flare triggering, magnetic complexity of active regions, such as that of $\delta$ spots, is more important than their size.

For some time, it has been known that stellar flares are somewhat different from the standard solar model. Cooler than our Sun, stars such as K and M dwarfs produce flares that seem both surprisingly energetic (flares on active M dwarfs are typically 10–1000 times as energetic as solar flares) and qualitatively different from solar flares, showing strong continuum or "white-light" emission, which resembles a 9000–10000 K blackbody superimposed over the quiet spectrum of the star (Walkowicz et al. 2011). Flares in G-type stars follow the same pattern seen in K and M stars, however the flares are always more energetic and more frequent (Maehara et al. 2015; Balona & Abedigamba 2016; Notsu et al. 2013).

In this paper, we carry out a study of stellar activity from the modelling of starspots, using the planetary transit mapping method. We investigate the differences in stellar parameters of two otherwise very similar stars with and without superflares, with the goal of inferring possible triggering mechanisms for superflares.

## 2 A CASE STUDY: KEPLER-210 AND KEPLER-411

For the study of stellar activity, we have chosen two stars with similar stellar and planetary system parameters. The two K-type stars (Table 1) are:

- Kepler-210 (KIC 7447200) is a K-type active star, with two transiting mini-Neptune like exoplanets. The Kepler-210 data are avail- able in the short-cadence format for the Q7-Q17 quarters, where we use the short-cadence data in Pre-search Data Conditioning (PDC- SAP) format for our analysis.

- Kepler-411 (KIC 11551692) is a K2V-type star exhibiting characteristics that indicate relatively strong magnetic activity (Sun et al.2019).

Four small exoplanets orbit the star, with three of them transiting. The Kepler-411 data are available in the short-cadence data for Q11-Q17. Here, we use the short-cadence data in PDCSAP format for our analysis.

⋆ E-mail: adesouza.astro@gmail.com

**Table 1.** Stellar parameters of Kepler-411 and Kepler-210.

| Stellar parameters | | |
|---|---|---|
| Parameter | Kepler-411 | Kepler-210 |
| Spectral Type | K2V[a] | K[c] |
| Radius [$R_\odot$] | $0.820 \pm 0.018$[a] | $0.85^c$ |
| Mass [$M_\odot$] | $0.87 \pm 0.04$[a] | $0.63^c$ |
| $T_{eff}$ [K] | $4832^b$ | $4559^d$ |
| Period [days] | $10.4 \pm 0.03$[a] | $12.33 \pm 0.05^c$ |

[a](Sun et al. 2019); [b](Gaia et al. 2018); [c](Ioannidis et al. 2014); [d]Exoplanet.eu.

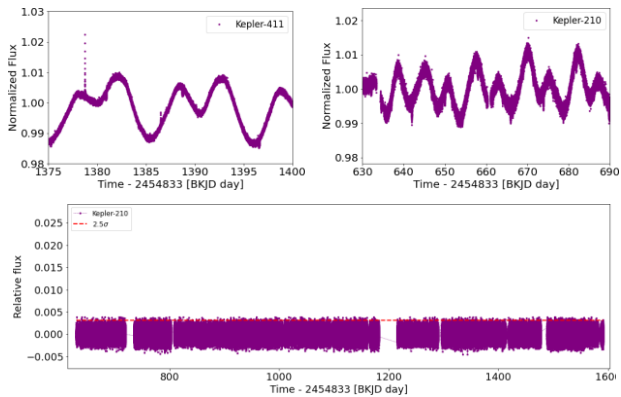

**Figure 1.** Modulation observed in the light curves due to starspots on the surface of the stars. **Top right panel:** Kepler-411 with the presence of superflares and **Top left panel:** Kepler-210. **Bottom panel:** Kepler-210 light curve normalized without corrections. To be considered a superflare, we adopted the requirement that 3 or more data points have to be above the 2.5 $\sigma$ limit (red dashed line).

The characteristics of the planets orbiting the two active stars are listed on Table 2. The planets size range from super-Earth to mini-Neptunes with radius of about 0.04 $R_{star}$. All exoplanets are in close orbits around their host star. These two stars, Kepler-411 and Kepler-210, were observed by the Kepler mission for about 600 days and 1000 days, respectively. The resulting light curves of both stars are shown in Figure 1, for Kepler-411 (top right panel) and Kepler-210 (top left panel).

## 3 ACTIVITY MODELLING

### 3.1 Physical Parameters of Starspots

We modelled the stellar spots on both stars using the Silva (2003) model, that was the first to propose to use an exoplanet as a probe to investigate the presence of spots on the surface of stars. From the spot occultation by a transiting planet, it is possible to physically characterize spots in solar-type stars (FGK) and M stars. This method requires high precision data, such as those obtained from the CoRoT, Kepler, and TESS space telescopes.

The Silva (2003) transit spot mapping model may be applied to light curves of stars with transiting exoplanets. First a 2D synthesized image of the star is constructed, considering limb darkening, and the light curve of the transit of the planet in its orbit is simulated. Also the model enables the inclusion of spots (dark) and faculae (bright regions) on the surface of the star. A simulated planetary transit exhibits small "bumps" in the light curve as the planet occults the dark spots on the stellar surface. Basically, the height of this small flux variation is a measure of the spot's intensity whereas the duration of this bump reflects the spot's size.

This model has been successfully applied to a few solar-type stars (Silva-Valio et al. 2010; Valio et al. 2017; Zaleski et al. 2019; Netto & Valio 2020; Araújo & Valio 2021a; Zaleski et al. 2022; Valio & Araújo 2022). Other authors have also applied the same method to other spotted stars (e.g. Sanchis-Ojeda & Winn 2011; Morris et al. 2017). This technique of spot transit mapping yields the physical parameters of the spots such as: size, intensity, temperature, and location (longitude and latitude) on the surface of the star. The temperature of the spot is obtained from the spot intensity fraction, $f = \frac{I_{spot}}{I_{star}}$ with respect to the star's central intensity, $I_{star}$, by assuming that both the star and the spot emit radiation as blackbodies, following Eq.2 of Silva-Valio et al. (2010). The spots on Kepler-411 have already been analysed by Araújo & Valio (2021a), whereas those on Kepler-210 are reported in Valio & Araújo (2022).

### 3.2 Identification of Superflares

We have carried out visual inspection of the light curves of Kepler-210, looking for impulsive brightness excess. We removed any indicator of data contamination according to Table 2–3 of Thompson et al. (2016), and then subtracted a high order polynomial so as to eliminate the spot modulation of the light curve with timescale of the stellar rotation period. The resulting light curve of Kepler-210 is plotted in bottom panel of Figure 1. For each quarter, candidate flares were identified in the light curve when three or more consecutive points were above the mean flux plus 2.5$\sigma$, where $\sigma$ is the standard deviation. As can be seen from the figure, no flare whatsoever can be identified in the light curve. On the other hand, a total of 65 superflares have been identified in Kepler-411 light curve, that have already been analysed by Araújo & Valio (2021b).

## 4 COMPARISON OF STARSPOTS ON KEPLER-411 AND KEPLER-210

For the first time, the stellar activity of two K-type stars observed by the Kepler telescope are compared. The goal is to understand the mechanisms that influence the occurrence of stellar superflares.

One of the stars, Kepler-411 (K2V-type), exhibits intense stellar activity with spots and superflares, whereas the other star, Kepler-210 (K-type) also has starspots albeit no superflare. In addition to being of similar spectral type, with about the same physical parameters (see Table 1), these stars display similar rotation period and planetary systems.

The spot transit mapping model allows for high spatial resolution of spot's size and location on the stellar disk, as well as temperature estimate. Since these stars are transited by more than one planet, information about the spotted surface of these stars is obtained for more than one latitude. Hopefully, the spots parameters will provide clues about the explosive activity, or lack thereof, in these stars.

Spot modelling of Kepler-411 has already been performed by Araújo & Valio (2021a) using the transits of the 3 exoplanets, where a total of 198 starspots were detected. The distributions of spot's size, intensity with respect to disk centre, and temperature are shown in Figure 2 (gray histograms), and their average parameters listed in Table 3.

The spots on the star Kepler-210 were also modelled by applying the same technique described in (Silva 2003). On Kepler-210, we were able to identify 107 starspots using the transits of the 2 exoplanets (Valio & Araújo 2022). Histograms of the spots parameters (radius, intensity, and temperature) are shown in Figure 2 for Kepler-210 in purple, whereas the mean values are also listed in Table 3.

**Table 2.** Planetary Physical Parameters of Kepler-411$^{a,b,c}$ and Kepler-210$^{d,e}$

| Parameter | Kepler-411b | Kepler-411c | Kepler-411d | Kepler-210b | Kepler-210c |
|---|---|---|---|---|---|
| Orbital Period [days] | $3.0051 \pm 0.00005$ | $7.834435 \pm 0.000002$ | $58.02 \pm 0.0002$ | $2.4532 \pm 0.0000001$ | $7.9725 \pm 0.000003$ |
| Planet Radius [$R_\oplus$] | $1.88 \pm 0.02$ | $3.27^{+0.011}_{-0.006}$ | $3.31 \pm 0.009$ | $3.09 \pm 0.21$ | $4.13 \pm 0.14$ |
| Planet Radius [$R_{star}$] | $0.024 \pm 0.002$ | $0.042 \pm 0.002$ | $0.040 \pm 0.002$ | $0.041 \pm 0.001$ | $0.055 \pm 0.002$ |
| Semi-Major Axis [au] | $0.049 \pm 0.0006$ | $0.080 \pm 0.001$ | $0.29 \pm 0.0004$ | $0.034 \pm 0.002$ | $0.071 \pm 0.001$ |

$^a$Wang et al. (2014) $^b$exoplanet.eu $^c$Araújo & Valio (2021a) $^d$Valio & Araújo (2022) $^e$Rowe et al. (2014).

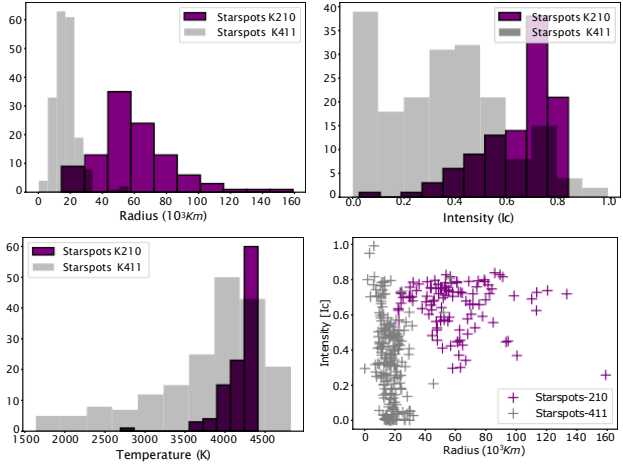

**Figure 2.** Histograms of spot physical parameters for Kepler-411 (gray) and Kepler-210 (purple): **Top left panel:** radius, **Top right panel:** intensity with respect to surrounding photospheric intensity, $I_c$, **Bottom left panel:** temperature. **Bottom right panel:** Intensity versus radius of starspots.

**Table 3.** Number of starspots and mean values of physical parameters.

| Parameter | Kepler-411$^a$ | Kepler-210$^b$ |
|---|---|---|
| Number of spots | 198 | 107 |
| Radius [$10^3$ km] | $17 \pm 7$ | $58 \pm 23$ |
| Intensity [$I_c$] | $0.35 \pm 0.24$ | $0.64 \pm 0.15$ |
| Temperature [K] | $3800 \pm 700$ | $4180 \pm 240$ |

$^a$Araújo & Valio (2021a) $^b$Valio & Araújo (2022).

The mean intensity of the spots can be converted to temperature by assuming that both the spot and the stellar photosphere emit as blackbodies (Valio et al. 2017). From the measured intensities, the average temperature of Kepler-411 spots was estimated to be $3800 \pm 700$ K, whereas those of Kepler-210 had mean temperatures of $4180 \pm 240$ K. Since Kepler-210 has a slightly cooler effective temperature ($T_{eff} = 4559$ K) than Kepler-411 ($T_{eff} = 4832$ K), it was expected that the spots on Kepler-210 would be a little bit cooler than those on Kepler-411. However, the ratio of spot average temperature to stellar effective temperature is $0.79 \pm 0.14$ for Kepler-411 and $0.88 \pm 0.07$ for Kepler-210.

Moreover, a surprising result found is the large difference in the size of the spots between the two stars. The mean radius of spots on Kepler-411 is $(17 \pm 7) \times 10^3$ km, whereas those on Kepler-210 are $(58\pm23) \times 10^3$ km, much larger, despite the fact that Kepler-210 stellar radius is 20% smaller than that of Kepler-411. All spots detected by the transit method in Kepler-410 and Kepler-210 are shown in the top panels of Figure 3, whereas the bottom panels represent the maps of spots in time (for each transit) as a function of spot longitude in the rotating frame of the star, taking into consideration the stellar differential rotation.

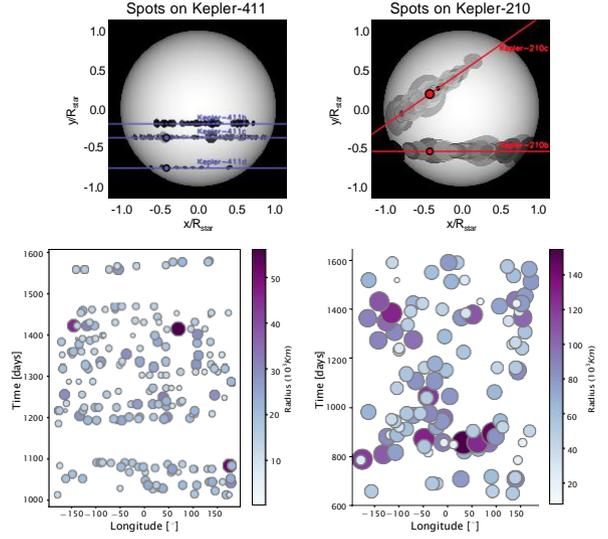

**Figure 3.** Distribution of starspots from planetary transits, following Eq. 2, 3 and 4 of Araújo & Valio (2021a). **Top left:** Transits of Kepler-411 planets b, c and d on the spotted star. **Top right:** Transits of planets Kepler-210b and Kepler-210c, which is in an oblique orbit Valio & Araújo (2022). Spot maps of longitude in time, depicting the spot size on the surface of the **Bottom left:** Kepler-411 and **Bottom right:** Kepler-210. The spots longitude is measured in a coordinate system that rotates with the star, where zero longitude is that of the midtransit time of the first transit of planet Kepler-411c and Kepler-210c, respectively. Notably, the surface area coverage of spots of Kepler-210 is much larger than that of Kepler-411.

Valio et al. (2020) analyzed the physical characteristics of over 32,000 sunspots during Solar Activity Cycle 23, such as area, intensity (or temperature), and magnitude of the extreme magnetic field. As a result, linear correlations were found between the logarithm of the area and the extreme magnetic field, between this magnetic field and spot intensity, as well as between temperature and the logarithm of the area. In the case of spots on Kepler-411 and Kepler-210, we found no evidence of correlation between starspot intensity and radius, as can be seen in the bottom right panel of Figure 2. We note that there may exist a degeneracy between these quantities, which is a result of the modeling.

## 5 DISCUSSION

Photometric modulations seen in stellar light curves, with periodicity of the stellar rotation, are usually associated with the presence of large spots that come and go into view as the star rotates. Several works claim that stars with larger spots produce large flares (Maehara et al. 2012, 2015; Notsu et al. 2013, 2015; Balona 2015; Yang et al. 2017). However, a few G-type stars, such as Kepler-17 (Valio et al. 2017) and the young star Kepler-63 (Netto & Valio 2020), with clear evidence of many spots on their surface produced no detectable superflares. For comparison, Kepler-17 spots have a mean radius of $(57 \pm 28) \times 10^3$ km and Kepler-63 of $(32 \pm 14) \times 10^3$ km. Another example is a K-type stars, KOI-883 (Zaleski et al. 2022), has spots with a mean radius of $(57 \pm 16) \times 10^3$ km and features only 2 superflares. As for M-type stars, Kepler-45 (Zaleski et al. 2020) has a mean radius of $(45 \pm 16) \times 10^3$ km and has not produced superflares, whereas Kepler-1651 with 95 superflares (Araújo et al. 2022) exhibits spots with a mean radius of $(13 \pm 5) \times 10^3$ km.

The discrepancy in spot radius on the stars Kepler-411 and Kepler-210 are clearly seen in Figure 3. Nevertheless, the mean values of spot sizes of Kepler-411 and Kepler-210 are within the range exhibited by the other stars. Here it is important to note that both the mean radius and the temperature, or intensity, are higher for the spots on Kepler-210. This becomes a complicating factor to explain the absence of superflares by Kepler-210. Studies of solar active regions show that larger active regions produce more intense X-ray flares and tend to have a larger spot area, as well as complex flares with greater morphological and magnetic complexity (Toriumi & Wang 2019).

In the solar case, some active regions produce many flares, while others are flare-quiet (Toriumi & Wang 2019). The authors of this review discuss in depth three factors that may be responsible for flaring: the active region's (1) size, (2) magnetic complexity, and (3) evolution. Albeit the influence of the active region size, a more important factor is the magnetic complexity such as that found in $\delta$ spots. As counterexample is the largest sunspot ever detected on the Sun (April 1947), with an area of 6136 MSH (or equivalent radius of $43.5 \times 10^3$ km), which never produced a flare and was thought to have a $\beta$ magnetic configuration (Toriumi & Wang 2019).

On the other hand, complex sunspot groups of mixed magnetic polarities, such as $\delta$ spots, associated to active regions are ideal locations for eruptive events such as flares and/or CMEs. From a study of X-ray flares more energetic than X1 ($10^{-4}$ W m$^{-2}$) GOES class, Sammis et al. (2000) report that 82% of these flares occur in $\delta$ spots, reaching 100% for flares larger than X4 ($4 \times 10^{-4}$ W m$^{-2}$), whereas only 24% of >X1 flares occur in big spots, with area larger than 1000 MHS (equivalent radius of $17.6 \times 10^3$ km).

Solar white-light flares are the analogs of stellar flares. Thus to understand the relationship between the area of active regions and flares, we analyzed the 50 more energetic solar white-light flares provided by Namekata et al. (2017). We searched for the corresponding solar active region, based on the location reported in their Table 1, to obtain the area of each active region. The energy of these 50 solar flares are plotted against the area of the corresponding active region where they occurred in Figure 4. As can be seen from the figure, no positive correlation is seen for these energetic white light solar flares. It is important to highlight the difficulty in studying white-light flares on the Sun. Only a few events were cataloged, mainly due to observational bias, since to detect a solar white-light flare, a high contrast observation is necessary.

Rubenstein & Schaefer (2000) proposed that superflares are caused by magnetic reconnection between fields of the primary star and a close-in Jovian planet. However, there is no evidence of this rela-tionship in the stars studied by Maehara et al. (2012, 2015). And it is certainly not the case for the planetary systems studied here, since all planets are mini-Neptunes or smaller.

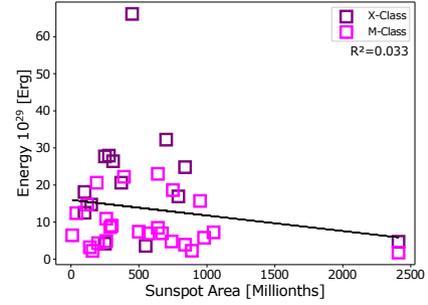

**Figure 4.** Energy of 50 energetic white-light solar flares (Namekata et al. 2017) of GOES classes X and M as a function of the sunspot area of the active region where they occurred.

Notsu et al. (2013) investigated the relationship between the energy and frequency of superflares and the rotation period of the star, taken as the periodicity of the light curve modulation. However, stars with relatively slower rotation rates can still produce flares that are as energetic as those of more rapidly rotating stars, although the average flare frequency is lower for slow rotators. For a sample of ultrafast rotating stars, Ramsay et al. (2020) detected that for a significant portion of stars, the number of flares appeared to drop significantly at periods <0.2 d. This is surprising, since this fast rotation should induce more intense magnetic fields, and thus present more flare activity. The two stars studied here have similar rotation periods of 10.4 and 12.33 days for Kepler-411 and Kepler-210, respectively (see Table 1).

The intense magnetic activity represented by the presence of spots on the stellar surface should be an indication of a higher probability of superflare occurrence. But recent work on starspot modelling, such as in Kepler-17 (Valio et al. 2017; Namekata et al. 2020), did not detect any superflares, albeit an extensive number of large starspots. Maehara et al. (2017) investigated solar-type stars with large starspots, but found no superflares. For Namekata et al. (2020) the explanation for the absence of superflares in these stars is the simple magnetic complexity of the spots, thought to be of type $\alpha$ or $\beta$, however direct observation of stellar magnetic configuration is not yet possible.

From Zeeman–Doppler imaging (ZDI), See et al. (2016) determined the magnetic field topology of solar-type stars. The authors found that active stars presented a dominant average toroidal field with large temporal variations, whereas the magnetic field of inactive stars was predominantly poloidal throughout their entire cycle. Böhm-Vitense (2007) argues that the difference in active cycle stars and inactive ones is the dynamo action at work. The stars in the active branch, with clear magnetic cycles have the dynamo mechanism acting in a shear layer near the stellar surface, whereas in inactive stars, the dynamo action takes place deep in the tachocline, the shear boundary layer between the radiative and convective zones.

## 6 CONCLUSION

To better understand stellar magnetic phenomena, we investigated possible relations between stellar superflares and the physical characteristics of starspots, by studying two K-type stars with similar physical parameters (mass, radius, $T_{eff}$, rotation). Araújo & Valio (2021a) found positive correlations between the distribution of starspots and the energy of superflares in Kepler-411. Here we investigated whether the same phenomena occurred in another K-type star, Kepler-210. The same starspot modelling and superflares identification methodology was applied, resulting that Kepler-210 exhibited large starspots, but no superflares were detected (Valio & Araújo 2022).

The analysis of spots on the stars Kepler-411 and Kepler-210 resulted in different characteristics, specially their size. Kepler-210 spots are larger than Kepler-411 spots, more than three times the radius. Also, albeit a small difference, the spots on Kepler-210 are slightly warmer than those on Kepler-411, when they should be colder due to the cooler effective temperature of Kepler-210 (4559 K). This reflects the fact that the average intensity ratio of the Kepler-210 spots is 0.64 when compared to the 0.35 intensity ratio of the Kepler-411 spots (with respect to the surrounding photosphere). In the Sun, the umbra of sunspots is always darker than the surrounding penumbra, which has a smaller magnetic field, with the minimum intensity of the penumbra being about 0.65 (Valio et al. 2020).

All other characteristics of the stellar parameters of the two stars are fairly similar, as well as the size and distance of the orbiting planets. Thus, our conclusion is that the difference in flare productivity may lie in the spot characteristics. However, what we found is contrary to previous reports in the literature, where a correlation between flare energy and spot area coverage of the star was found. However, the method to infer the spot area in previous works is from the out–of–transit light curve modulation, whereas the method used in this work (spot occultation by a transiting planet) is more precise.

One explanation is that albeit large, the spots of Kepler-210 are of simple magnetic complexity ($\alpha$ or $\beta$), whereas those of Kepler-411 are of type $\delta$, which may be more compact than $\alpha$ or $\beta$ spots. Moreover, the smaller spot contrast (or warmer temperatures) of Kepler-210 may also be an indication of less intense magnetic fields. In the Sun, spots with intense magnetic fields are usually darker than the ones with smaller fields (Valio et al. 2020). This reinforces the idea of simpler magnetic complexity of the Kepler-210 spots.

A $\delta$-spot can be formed in three different ways (Toriumi & Wang 2019), all of which involves emergence of closely spaced different polarity spots, which invokes a dynamic flux tube emergence. Thus, the flare productive spots of Kepler-411 may be the result of a near surface dynamo action that produces a predominantly toroidal field, whereas a deeper and slower dynamo is responsible for the larger and simpler $\beta$ spots on Kepler-210, similar to the largest spot observed on the Sun that never produced a flare. See et al. (2016) find that the transition from stars with dominantly poloidal fields to stars with mainly toroidal fields occurs at a rotation of ~12 days. We note that Kepler-411 has a rotation period of 10.4 days, whereas Kepler-210 rotates with a period of 12.33 days.

Therefore, maybe large starspot area is not the main factor causing superflares. Our analysis showed that the size of the spot (or active region) is not enough to generate superflares, but rather the magnetic complexity such as that found in $\delta$ sunspots, is more important. Results from Doyle (2022) corroborate our findings, that is, the magnetic field strength may not be the answer to the lack of flaring activity of fast rotating stars, perhaps supersaturation or magnetic field configuration play an important role.

We note that the main result presented here, that the size of active region is not decisive in flare occurrence, was only possible because of the spatially resolved analysis of starspots through transit mapping. Unfortunately, the current data obtained with the current space missions is not able to decipher the magnetic complexity of starspots, hampering the explanation of why some spotted stars produce superflares and others do not.

Understanding stellar magnetic activity is important due to the di-rect impact of superflares onto the atmosphere of exoplanets. Moreover, this will also shed light onto the probability estimates of intense flares occurring on the present Sun.

## ACKNOWLEDGEMENTS AND DATA AVAILABILITY

We are grateful to the anonymous referee for the suggestions that helped improve this work. The authors acknowledge partial financial support from FAPESP grant #2013/10559-5, CNPq grant #150817/2022-3. The reduced data analyzed in this article is available upon request.